\documentclass[
    ,final            
  ]
  {aipproc}

\layoutstyle{6x9}
\usepackage{wrapfig,rotating}
\usepackage{amssymb,amsmath,array}
\usepackage{mathrsfs}

\newcommand{\bea}{\begin{eqnarray}}
\newcommand{\eea}{\end{eqnarray}}
\newcommand{\mA}{\mathscr{A}_{TT}}

\begin{document}

\title{Double Transverse-Spin Asymmetries for\\ 
Small-$Q_T$ Drell-Yan Pair Production\\ 
in $p\bar{p}$ Collisions}


\author{Hiroyuki Kawamura}{
  address={Department of Mathematical Sciences, University of Liverpool, 
Liverpool L69 3BX, UK}
}

\author{Kazuhiro Tanaka}{
  address={Department of Physics, Juntendo University, Inba, Chiba 270-1695, Japan}
}

\begin{abstract}
We discuss
the double-spin asymmetries 
in transversely polarized Drell-Yan process, calculating 
all-order gluon resummation corrections up to the next-to-leading logarithmic accuracy.
This resummation is 
relevant when the transverse-momentum $Q_T$
of the produced lepton pair is small, and 
reproduces the (fixed-order) next-to-leading QCD corrections upon integrating over $Q_T$.
The resummation corrections in $p\bar{p}$-collision behave 
differently 
compared with 
$pp$-collision cases,
and 
are small 
at the kinematics 
in the proposed GSI experiments.
This 
fact 
allows us to predict large value of 
the double-spin asymmetries 
at GSI,
using recent
empirical information on the transversity.
\end{abstract}

\maketitle

The double-spin asymmetry in Drell-Yan (DY) process 
with transversely-polarized protons and antiprotons,
$p^{\uparrow}\bar{p}^{\uparrow}
\rightarrow l^+l^-X$, for azimuthal angle $\phi$ of a lepton
measured in the rest frame of the dilepton $l^+l^-$ with invariant mass $Q$ and rapidity $y$,
is given by 
\begin{equation}
A_{TT}  = 
\frac{d\sigma^{\uparrow  \uparrow}/d\omega - d\sigma^{\uparrow  \downarrow}/d\omega}
{d\sigma^{\uparrow  \uparrow}/d\omega + d\sigma^{\uparrow  \downarrow}/d\omega}
 \equiv \frac{\Delta _T d\sigma /d\omega}{d\sigma /d\omega}
= \frac{\cos (2\phi )}{2}\frac{\sum_q e_q^2 \delta q(x_1 , Q^2)\delta q(x_2 , Q^2) +\cdots}
{\sum_q e_q^2 q(x_1 ,Q^2)q(x_2 ,Q^2) +\cdots}\ ,
\label{eq:1}
\end{equation}
where $d\omega \equiv dQ^2 dy d\phi$, the summation is over all 
quark and antiquark flavors
with $\delta q(x, Q^2)$ and $q(x, Q^2)$ being the transversity and unpolarized 
quark-distributions inside a proton,
and the ellipses stand for the corrections of next-to-leading order (NLO) and higher
in QCD perturbation theory.
The scaling variables $x_{1,2}$
represent the momentum fractions associated with the partons
annihilating via the DY mechanism, such that 
$Q^2=(x_1 P_1 + x_2 P_2)^2 = x_1 x_2 S$ and $y=( 1/2)\ln (x_1 /x_2)$,
where
$S=(P_1 +P_2)^2$ is the CM energy squared of $p^{\uparrow}\bar{p}^{\uparrow}$.
In the proposed polarization experiments at GSI~\cite{Barone:2005pu},
moderate energies, $30 \lesssim S \lesssim 200$~GeV$^2$,
allow us to measure (\ref{eq:1}) for $0.2 \lesssim Q/\sqrt{S}\lesssim 0.7$,
and probe the product of the two quark-transversities
in the ``valence region''.
Recently, QCD corrections 
in (\ref{eq:1})
at GSI kinematics have been studied:
the NLO ($O(\alpha_s)$) 
corrections~\cite{Barone} as well as the
higher order ones
in the framework of threshold resummation~\cite{SSVY:05}
are rather 
small, so that the LO value of 
$A_{TT}$, which turns out to be large, is rather robust.

When the transverse momentum $Q_T$ of the final $l^+l^-$ 
is also observed,
we obtain the new double-spin asymmetry 
as the ratio of the $Q_T$-differential cross sections, 
$\mA (Q_T) \equiv [\Delta _T d\sigma/d\omega dQ_T]/[d\sigma /d\omega dQ_T]$.
The bulk of $l^+l^-$ pair is produced at small $Q_T \ll Q$,
where the cross sections $(\Delta _T ) d\sigma/d\omega dQ_T$ 
receive large perturbative corrections 
with logarithms $\ln ( Q^2 / Q_T^2 )$ multiplying $\alpha_s$
at each order, by the recoil 
from gluon radiations, and those
have to be treated
in an all-order resummation 
in QCD~\cite{CSS}.
The corresponding ``$Q_T$-resummation'' has formally some resemblance 
to the threshold resummation~\cite{SSVY:05}, but embodies 
contributions 
from
different ``edge region''
of phase space.
The $Q_T$-resummation for $\Delta _T  d\sigma/d\omega dQ_T$ 
has been derived recently~\cite{KKST:06}, 
summing the corresponding 
large logarithms up to next-to-leading logarithmic (NLL) accuracy. 
Combined with that for $d\sigma/d\omega dQ_T$~\cite{CSS}, we get~\cite{KKT:08}
($b_0\equiv 2e^{-\gamma_E}$ with $\gamma_E$ the Euler constant)
\begin{equation}
\mA(Q_T) =\frac{\cos (2\phi )}{2}
\frac{\int d^2 b\ e^{i \mathbf{b} \cdot \mathbf{Q}_T} 
e^{S(b,Q)} 
\sum\nolimits_q  e_q^2 
\delta q (x_1 , b_0^2/b^2 ) 
\delta q(x_2 , b_0^2/b^2 ) +\cdots}
{\int d^2 b\ e^{i \mathbf{b} \cdot \mathbf{Q}_T} 
e^{S(b,Q)} 
 \sum\nolimits_q  e_q^2 
q (x_1 , b_0^2/b^2 ) 
q(x_2 ,b_0^2/b^2 ) +\cdots
},
\label{eq:2}
\end{equation}
where 
the numerator and denominator are, respectively, reorganized
in the impact parameter $b$ space in terms of 
the Sudakov factor $e^{S(b,Q)}$ resumming soft and flavor-conserving collinear
radiation, while the ellipses involve the remaining contributions of the $O (\alpha_s)$
collinear radiation, which can be absorbed into the exhibited terms
as 
$\delta q \rightarrow \Delta_T C_{qq}\otimes  \delta q$,
$q \rightarrow C_{qq} \otimes  q + C_{qg}\otimes g$ using the corresponding coefficient 
functions $(\Delta_T) C_{ij}$; 
note that there is no ``chiral-odd'' gluon distribution 
to participate in the numerator
of (\ref{eq:2}) as well as (\ref{eq:1}).
Using 
{\em universal} Sudakov exponent $S(b,Q)$ with the first nonleading 
anomalous dimensions in (\ref{eq:2}), 
the first three towers of large logarithmic contributions
to the cross sections,
$\alpha_s^n\ln^m(Q^2/Q_T^2)/Q_T^2$ ($m=2n-1, 2n-2, 2n-3$),
are resummed to all orders in $\alpha_s$,
yielding the NLL resummation.
In addition to these resummed components relevant for small $Q_T$,
the ellipses in (\ref{eq:2}) also involve  
the other terms of the fixed-order $\alpha_s$, which treat the LO processes in the large
$Q_T$ region, so that (\ref{eq:2}) 
is the ratio of the NLL+LO polarized and unpolarized cross sections.
We include a Gaussian smearing 
as usually
as $S(b,Q) \rightarrow S(b,Q) -g_{NP}b^2$,
corresponding to intrinsic transverse momentum of partons inside proton.
The integrations of the NLL+LO cross sections
$\Delta_T  d\sigma/d\omega dQ_T$ and $d\sigma/d\omega dQ_T$ over $Q_T$
coincide, respectively, with 
the (fixed-order) NLO cross sections $\Delta_T d\sigma/d\omega$ and
$d\sigma/d\omega$,
associated with $A_{TT}$ of~(\ref{eq:1}) \cite{KKST:06,KKT:08}.

\begin{figure}[!t]
\includegraphics[height=4.5cm]{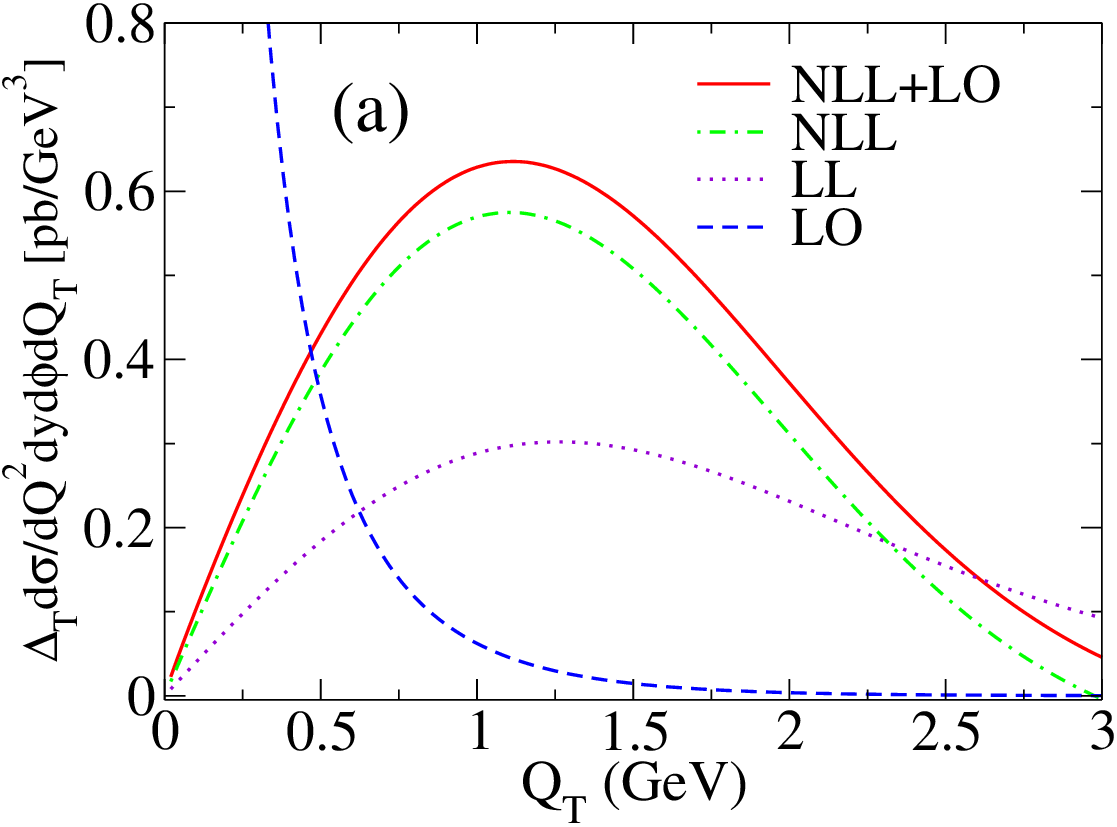}~~~~~~~~~~
\includegraphics[height=4.5cm]{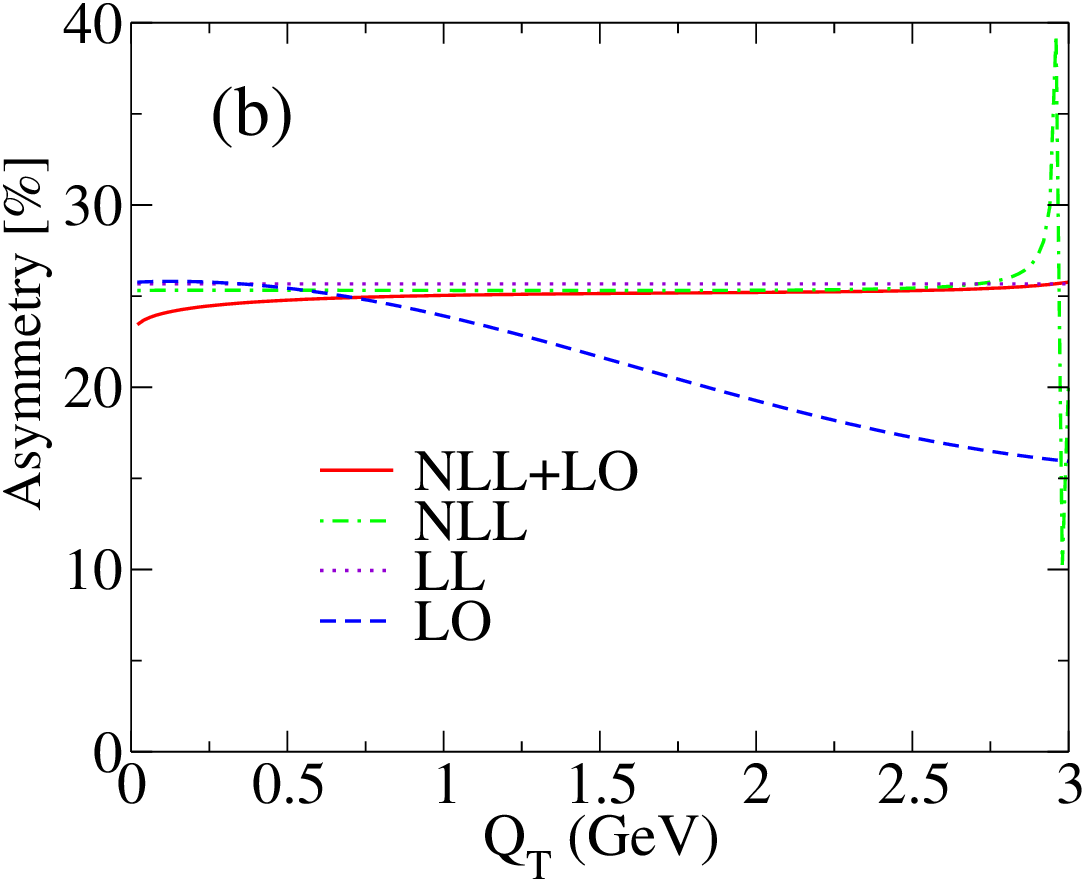}
\caption{(a) 
$\Delta_Td\sigma/d\omega dQ_T$
(b) $\mA(Q_T)$ 
with GSI kinematics, $S=210$~GeV$^2$, $Q=4$~GeV,
$y=\phi=0$, and with $g_{NP}=0.5$~GeV$^2$, using the NLO 
transversity distributions  
that correspond to the Soffer bound.}
\label{fig:2}
\end{figure}

The resummation 
makes $1/b \sim Q_T$ the relevant scale in (\ref{eq:2}), 
in contrast to $Q$ in (\ref{eq:1}).  
Figure~\ref{fig:2} shows~\cite{KKT:08} the numerical evaluation of (\ref{eq:2}), 
as well as of 
$\Delta_T  d\sigma/d\omega dQ_T$
associated with its numerator, at GSI kinematics 
using 
the NLO transversities 
that saturate the Soffer bound, $2\delta q(x,\mu^2)\le q(x,\mu^2)+\Delta q(x,\mu^2)$, 
at a low scale $\mu$ with $\Delta q$ the helicity distribution.
The NLL resummed component dominates $\Delta_T  d\sigma/d\omega dQ_T$
in small and moderate $Q_T$ region, and similarly for $d\sigma/d\omega dQ_T$ 
reflecting universality of the large Sudakov effects,
which leads to almost constant $\mA(Q_T)$, in particular, 
with even flatter behavior than the corresponding asymmetry~\cite{KKST:06} for the $pp$-collision case.
Remarkably, $\mA(Q_T)$ at NLL+LO has almost the 
same value as that at LL; this is in contrast to the $pp$ case where
the resummation at higher level
enhances the asymmetry~\cite{KKST:06}.
We note that $\mA(Q_T)$ at LL is given by (\ref{eq:2}) 
omitting all nonleading corrections, i.e., omitting the ellipses,
replacing $S(b,Q)$ by that at the LL level, and replacing the scale of the parton distributions 
as $b_0^2/b^2 \rightarrow Q^2$,
so that the result coincides with $A_{TT}$ at LO (see (\ref{eq:1})).
Therefore, at GSI, both $\mA(Q_T)$ and $A_{TT}$
are quite stable when including the 
QCD
(resummation and fixed-order)
corrections, with $\mA(Q_T) \simeq \mA(0)$, and
\begin{equation}
\mA(Q_T) \simeq A_{TT}.
\label{eq:4}
\end{equation}

To clarify the relevant mechanism, 
$\mA(Q_T) \simeq \mA(0)$
allows us to consider the $Q_T \rightarrow 0$ limit of (\ref{eq:2}):
for $Q_T \approx 0$,
the $b$ integral 
is controlled by a saddle point $b=b_{SP}$, which has the same
value between the numerator and denominator in (\ref{eq:2}) 
such that~\cite{KKST:06,KKT:08}
\begin{equation}
\mA(0)\simeq \frac{\cos (2\phi )}{2}\frac{\sum_q e_q^2 \delta q(x_1 , b_0^2/b_{SP}^2 )
\delta q(x_2 , b_0^2/b_{SP}^2 )}
{\sum_q e_q^2 q(x_1 ,b_0^2/b_{SP}^2 ) q(x_2 ,b_0^2/b_{SP}^2 )}\ ,
\label{eq:3}
\end{equation}
omitting the small corrections
from the LO components involved in the ellipses in (\ref{eq:2}).
This saddle-point evaluation is exact at
NLL accuracy; in particular, the $O(\alpha_s)$ contributions
from the coefficients $(\Delta_T) C_{ij}$, e.g., $C_{qg}\otimes g$ associated with gluon distribution, 
completely decouple as $Q_T \rightarrow 0$ (see \cite{CSS,KKST:06}).
The simple form of (\ref{eq:3}) is reminiscent of $A_{TT}$ of (\ref{eq:1}) at LO,
but is different from the latter in the 
unconventional scale $b_0^2/b_{SP}^2$; 
the actual position of the saddle point implies
$b_0/b_{SP}\simeq 1$~GeV, 
irrespective of the values of $Q$ and $g_{NP}$~\cite{KKST:06,KKT:08}.
In the valence region 
relevant for GSI kinematics, the $u$-quark contribution dominates in (\ref{eq:3})
and (\ref{eq:1}), so that these asymmetries are controlled by the ratio,
$\delta u(x_{1,2}, \mu^2)/u(x_{1,2},\mu^2)$
with $\mu^2 = b_0^2/b_{SP}^2$ and $Q^2$, 
respectively. Actually
the scale dependence in this ratio
almost cancels between the numerator and denominator 
as $\delta u(x,  b_0^2/b_{SP}^2)/u(x,  b_0^2/b_{SP}^2)
\simeq \delta u(x, Q^2)/u(x, Q^2)$
(see Fig.~3 in \cite{KKT:08}); 
this implies 
(\ref{eq:4})
at GSI.
Note that this is not the case 
for $pp$ collisions 
because of very different behavior of the sea-quark components under the evolution
between transversity and unpolarized distributions~\cite{KKST:06}; indeed $\mA(Q_T) > A_{TT}$ 
at RHIC and J-PARC~\cite{KKST:06}.
A similar logic applied to (\ref{eq:2})
also explains why $\mA(Q_T)$
at GSI
are flatter 
than in $pp$ collisions.

Another consequence of the similar logic 
is that
$\delta u(x_{1,2}, 1~{\rm GeV}^2)/u(x_{1,2}, 1~{\rm GeV}^2)$ 
as a function of $x_{1,2}$ directly determines the $Q$- as well as $S$-dependence 
of the value of (\ref{eq:4}) at GSI, 
through $x_{1,2}=(Q/\sqrt{S})e^{\pm y}$. 
In Fig.~{\ref{fig:1}, using the NLO transversity distributions
corresponding to the Soffer bound,
the symbols ``$\bigtriangleup$''
plot $\mA(Q_T\simeq 1~{\rm GeV})$
of (\ref{eq:2}) at NLL+LO~\cite{KKT:08}.
\begin{figure}[!t]
\includegraphics[width=0.55\textheight,clip]{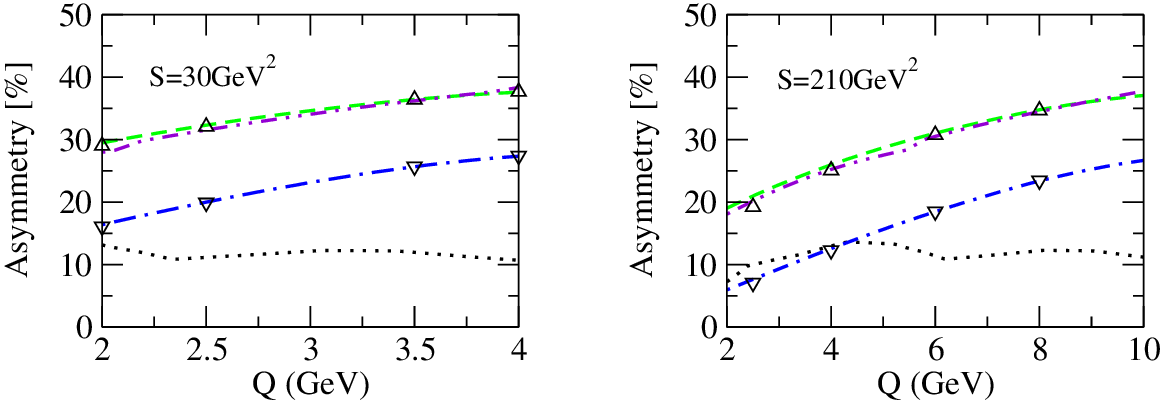}
\caption{
The double transverse-spin asymmetries at GSI as functions of $Q$ 
with $y=\phi=0$.
}
\label{fig:1}
\end{figure}
The dashed curve draws the result using 
(\ref{eq:3});
this simple 
formula indeed works well. 
Also plotted by the two-dot-dashed curve
is $A_{TT}$ of (\ref{eq:1}) at LO
with the transversities corresponding to the Soffer bound at LO level,
to demonstrate (\ref{eq:4}).
The $Q$- and $S$-dependence of these results 
reflect that
the ratio $\delta u(x, 1~{\rm GeV}^2)/u(x, 1~{\rm GeV}^2)$ 
is an increasing function 
of $x$.
These results using the Soffer bound 
show the ``maximally possible'' asymmetry, i.e., 
optimistic 
estimate.
A more realistic estimate of (\ref{eq:2}) (with $Q_T \simeq 1$~GeV) and (\ref{eq:3}) 
is shown \cite{KKT:08} in Fig.~{\ref{fig:1} by the symbols
``$\bigtriangledown$'' and the dot-dashed curve, respectively,
with the NLO transversity distributions
assuming $\delta q(x, \mu^2)=\Delta q(x, \mu^2)$ at a low scale $\mu$,
as 
suggested by 
nucleon models and favored
by the results of empirical fit for transversity \cite{Anselmino:07}.
The new estimate gives 
smaller asymmetries
compared with the Soffer bound results,
but still yields rather large asymmetries~\cite{KKT:08}.
Based on (\ref{eq:4}), these results in Fig.~{\ref{fig:1} may be considered as estimate of $A_{TT}$ of (\ref{eq:1}).

At present, empirical information of transversity is based on
the LO global fit, using
the semi-inclusive DIS
data and 
assuming that the antiquark transversities 
vanish, $\delta \bar{q}(x)=0$,
so that the corresponding LO
parameterization is
available only for $u$ and $d$ quarks~\cite{Anselmino:07}.
Fortunately, however, the dominance of the $u$-quark contribution
in the GSI kinematics allows quantitative evaluation of $A_{TT}$ at LO
using only this empirical information~\cite{KKT:08}:
the upper limit of the error band
for the $u$- and $d$-quark transversities
obtained by the global fit~\cite{Anselmino:07}
yields the ``upper bound'' of $A_{TT}$ shown
by the dotted curve in Fig.~{\ref{fig:1};
using (\ref{eq:4}), this result may be considered also as 
estimate 
of $\mA(Q_T)$.
In the small $Q$ region, our full NLL+LO result of $\mA(Q_T\simeq 1~{\rm GeV})$,
shown by ``$\bigtriangledown$'',  can be consistent
with estimate using the empirical LO transversity, but these results have rather 
different behavior
for increasing $Q$, 
because the $u$-quark transversity for the former
lies slightly outside the 
error band of the global 
fit for $x \gtrsim 0.3$~\cite{KKT:08}.
Thus, 
in the large asymmetries to be observed at GSI, 
the behavior 
of $\mA(Q_T)$, $A_{TT}$
as functions of $Q$
will allow us to determine
the detailed shape of transversity distributions.

\begin{theacknowledgments}
This work was supported by the Grant-in-Aid for Scientific Research 
No.~B-19340063. 
\end{theacknowledgments}


\end{document}